# A Session based Multiple Image Hiding Technique using DWT and DCT

Tanmay Bhattacharya

Asst. Professor Dept. of IT
JIS College of Engineering
Kalyani, West Bengal, India

Nilanjan Dey

Asst. Professor Dept. of IT
JIS College of Engineering
Kalyani, West Bengal, India

S. R. Bhadra Chaudhuri

Professor, Dept of E&TCE
Bengal Engineering & Science University
Shibpur Howrah, West Bengal, India

## ABSTRACT

This work proposes Steganographic technique for hiding multiple images in a color image based on DWT and DCT. The cover image is decomposed into three separate color planes namely R, G and B. Individual planes are decomposed into subbands using DWT. DCT is applied in HH component of each plane. Secret images are dispersed among the selected DCT coefficients using a pseudo random sequence and a Session key. Secret images are extracted using the session key and the size of the images from the planer decomposed stego image. In this approach the stego image generated is of acceptable level of imperceptibility and distortion compared to the cover image and the overall security is high.

## General Terms

Steganography.

## Keywords

DCT, Session Based Key, Pseudo Random Sequence, RGB Color planes.

## 1. INTRODUCTION

*Steganography* [1, 2, 3] is the process of hiding of a secret message within an ordinary message and extracting it at its destination. Anyone else viewing the message will fail to know it contains secret/encrypted data. The word comes from the Greek word "*steganos*" meaning "covered" and "*graphei*" meaning "writing".

LSB [4] insertion is a very simple and common approach to embedding information in an image in spatial domain. The limitation of this approach is vulnerable to every slight image manipulation. Converting image from one format to another format and back could destroy information secret in LSBs. Stego-images can be easily detected by statistical analysis like histogram analysis. This technique involves replacing N least significant bit of each pixel of a container image with the data of a secret message. Stego-image gets destroyed as N increases. Data hiding can also be done in the frequency domain. Cover Image is transformed using conventional transformation like DFT, DCT [5, 6], DWT [7, 8, 9] etc. Secret information is embedded in the less significant frequency components of cover image. The advantage of using frequency domain Steganography is that it is very secure, hard to detect, flexible and has different techniques for manipulation of DCT coefficients values.

## 2. DISCRETE WAVELET TRANSFORMATION

The wavelet transform describes a multi-resolution decomposition process in terms of expansion of an image onto a set of wavelet basis functions. Discrete Wavelet Transformation has its own excellent space frequency localization property. Applying DWT in 2D images corresponds to 2D filter image processing in each dimension. The input image is divided into 4 non-overlapping multi-resolution sub-bands by the filters, namely LL1 (Approximation coefficients), LH1 (vertical details), HL1 (horizontal details) and HH1 (diagonal details). The sub-band (LL1) is processed further to obtain the next coarser scale of wavelet coefficients, until some final scale "N" is reached. When "N" is reached, we'll have 3N+1 sub-bands consisting of the multi-resolution sub-bands (LLN) and (LHX), (HLX) and (HHX) where "X" ranges from 1 until "N". Generally most of the Image energy is stored in these sub-bands.

| $LL_3$ | $HL_3$ | $HL_2$ | $HL_1$ |
|---|---|---|---|
| $LH_3$ | $HH_3$ | | |
| $LH_2$ | | $HH_2$ | |
| $LH_1$ | | | $HH_1$ |

**Fig.1 Three phase decomposition using DWT.**

The Haar wavelet is also the simplest possible wavelet. Haar wavelet is not continuous, and therefore not differentiable. This property can, however, be an advantage for the analysis of signals with sudden transitions.

## 3. DISCRETE COSINE RANSFORMATION

Discrete Cosine Transform (DCT) [10,11] separate the image into parts (or spectral sub-bands) of differing importance (with respect to the image's visual quality). The DCT transforms a signal or image from the spatial domain to the frequency domain.

The general equation for a 2D (*N* by *M* image) DCT is defined by the following equation:

$$C(u,v) = \alpha(u)\alpha(v) \sum_{x=0}^{N-1}\sum_{y=0}^{N-1} f(x,y) \cos\frac{\pi(2x+1)u}{2N} \cos\frac{\pi(2y+1)v}{2N}$$

where

$$\alpha(u) = \sqrt{\frac{1}{N}} \quad \text{for} \quad u=0;$$

$$\alpha(u) = \sqrt{\frac{2}{N}} \quad \text{for} \quad u=1,2,3,\ldots, N-1;$$

and the corresponding *inverse* 2D DCT transform is as follows

$$f(x,y) = \sum_{u=0}^{N-1}\sum_{v=0}^{N-1} \alpha(u)\alpha(v) C(u,v) \cos\frac{\pi(2x+1)u}{2N} \cos\frac{\pi(2y+1)v}{2N}$$

The basic operation of the DCT is as follows:

- The input image is N by N matrix.
- f(x,y) is the intensity of the pixel in row x and column y.
- C(u,v) is the DCT coefficient in row k1 and column k2 of the DCT matrix.
- For most images, much of the signal energy lies at low frequencies; these appear in the upper left corner of the DCT.
- The lower right values represent higher frequencies and are of less visual importance.
- The DCT input is an 8 by 8 array of integers. This array contains each pixel's gray scale level.
- 8 bit pixels have levels from 0 to 255.
- The output array of DCT coefficients contains integers, these can range from - 1024 to 1023.

## 4. PROPOSED ALGORITHM
### 4.1 Secret Image Hiding

1. Cover colour image is decomposed into three colour planes (R, G and B).
2. Each colour plane is decomposed into four sub bands using DWT.
3. DCT is applied in HH band of each plane.
4. Information of three secret binary images are dispersed separately into selected high frequency components of each colour plane (in frequency domain) using a session based pseudo random 2D sequence.
5. After embedding bits of the secret images into three colour planes inverse transformations are applied to get back the planes in the spatial form.
6. Finally, three colour planes are combined to generate the colour stego image.

### 4.2 Secret Image Extraction

1. Stego colour image is decomposed into three colour planes (R, G and B).
2. DWT is applied in each plane to decompose into four sub-bands.
3. HH sub-bands of each plane are transformed using DCT.
4. Three secret images are extracted from the DCT coefficients of colour planes using the session based key and the size of the secret images which are known to the intended receiver only.

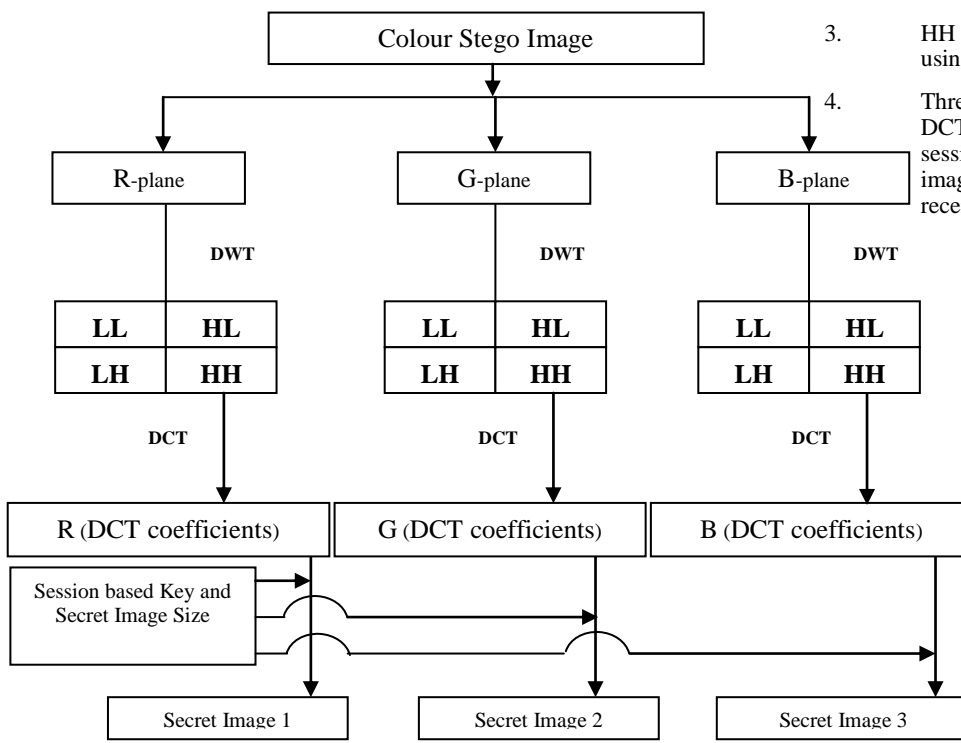

**Figure2. Image Extraction Process**

## 5. EXPLANATION OF THE ALGORITHM

### 5.1 Secret Image hiding procedure

The Cover Image (Color) is decomposed into three color planes namely Red, Green and Blue. Each color planes are decomposed into four sub-bands (LL, LH, HL, HH) using DWT.

DCT is applied in each HH band separately to get corresponding DCT coefficients.

Three secret binary images are converted into three 1-D vectors. Selected high frequency components of each plane are modified according to the bits of the individual image vector and the session based 2D pseudo random sequence.

Inverse transformation (IDCT) is applied to the Modified HH bands. Than IDWT is applied to combine four sub-bands including the modified HH sub-bands.

Three planes are then combined to generate the final color stego image.

### 5.2 Secret Image Extraction procedure

The color stego image is decomposed into RGB planes and DWT is applied to get HH band of each plane. Then DCT is applied to the HH bands separately. Extraction algorithm along with the session based key and the secret image size is used to recover three secret images from the color planes of the stego image.

## 6. RESULTS

Results of the proposed algorithm are shown below (Fig. 3 to Fig. 8).

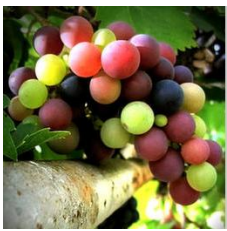
Figure3. Cover Image

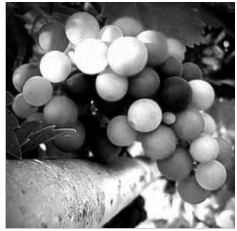
Figure 4. Cover Image(R Plane)

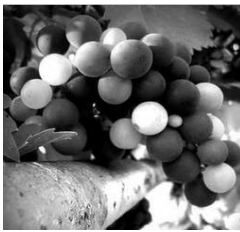
Figure 5. Cover Image (G Plane)

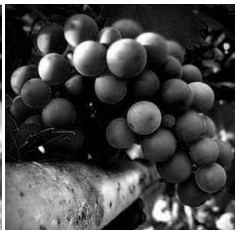
Figure 6. Cover Image (B Plane)

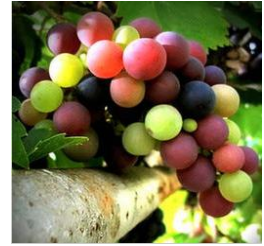
Figure 7. Stego Image

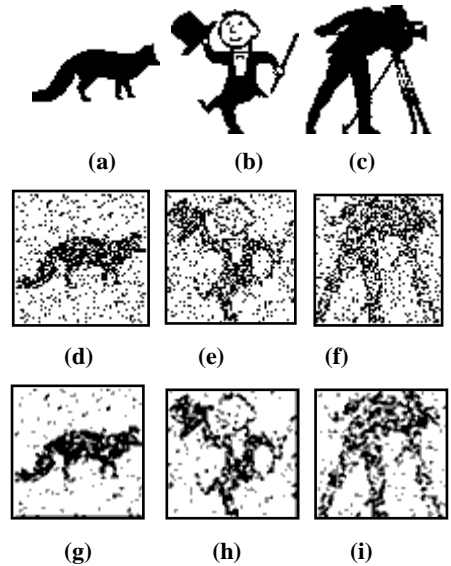
Figure 8. (a) (b) (c) Secret Images,(d) (e) (f) Extracted Secret Images,(g) (h) (i) Extracted Secret Images (Appling Image Filter)

### 6.1 Peak Signal to Noise Ratio (PSNR)

It measures the quality of a stego image. This is basically a performance metric and is used to determine perceptual transparency of the stego image with respect to host image:

$$PSNR = \frac{MN \max_{x,y} P_{x,y}^2}{\sum_{x,y}(P_{x,y} - \overline{P}_{x,y})^2}$$

Where, M and N are number of rows and columns in the input image,

$P_{x,y}$ is the original image and

$\overline{P}_{x,y}$ is the Stego Image.

PSNR between Cover Image and Stego Image is 40.1141 shown in Table1.

**TABLE 1**

| Cover Image | PSNR |
|---|---|
| vs. Stego Image | 40.1141 |

## 7. CONCLUSION

In this approach embedding is done randomly in the frequency domain. So, it will be difficult to detect the existence of the secret image using conventional steganalysis methods. Proposed approach gives satisfactory PSNR value to establish the robustness of the work. Since only selected high frequency components are modified for the hiding method, therefore there must be a constraint on the secret image size.

## 8. REFERENCES

[1] N. F. Johnson and S. Katzenbeisser," .*A survey of steganographic techniques*", in S. Katzenbeisser and F. Peticolas (Eds.): *Information Hiding*, pp.43-78. Artech House, Norwood, MA, 2000.

[2] Lou, D. C. and Liu, J. L. 2002. *"Steganography Method for Secure Communications"*. *Elsevier Science on Computers& Security*, 21, 5: 449-460.

[3] J. Fridrich and M. Goljan, .Practical steganalysis of digital images-state of the art., *Proc. SPIE Photonics West, Vol. 4675*, pp. 1-13, San Jose, California, Jan. 2002.

[4] Chan, C. K. and Cheng, L. M. 2003. Hiding data in image by simple LSB substitution. *Pattern Recognition*, 37:469-474.

[5] Iwata, M., Miyake, K., and Shiozaki, A. 2004. *"Digital Steganography Utilizing Features of JPEG Images",* IEICE Transfusion Fundamentals, E87-A, 4:929-936.

[6] Blossom Kaur, Amandeep Kaur, Jasdeep Singh, "*Steganographic Approach for Hiding Image in DCT Domain",* International Journal of Advances in Engineering & Technology, July 2011.

[7] Po-Yueh Chen* and Hung-Ju Lin, "*A DWT Based Approach for Image Steganography",* International Journal of Applied Science and Engineering 2006. 4, 3: 275-290

[8] Ali Al-Ataby and Fawzi Al-Naima, "*A Modified High Capacity Image Steganography Technique Based on Wavelet Transform",* The International Arab Journal of Information Technology, Vol.7, No. 4, October 2010

[9] Tanmay Bhattacharya, Nilanjan Dey,S. R. Bhadra Chaudhuri," A Novel Session Based Dual Steganographic Technique Using DWT and Spread Spectrum", International Journal of Modern Engineering Research , Vol.1, Issue1, pp- 157-161

[10] Ajit Danti, Preethi Acharya," Randomized Embedding Scheme Based on DCT Coefficients for Image Steganography" IJCA Special Issue on "Recent Trends in Image Processing and Pattern Recognition" RTIPPR, 2010.

[11] K B Shiva Kumar, K B Raja,R K Chhotaray, Sabyasachi Pattanaik," Bit Length Replacement Steganography based on DCT Coefficients", International Journal of Engineering Science and Technology Vol. 2(8), 2010, 3561-3570.